\documentclass{article}
\usepackage[margin=1in]{geometry}
\usepackage{cite}
\usepackage{amsmath,amssymb,amsfonts}
\usepackage{algorithmic}
\usepackage{graphicx}
\usepackage{textcomp}
\usepackage{xcolor}

\usepackage{tikz}
\usetikzlibrary{positioning}
\usetikzlibrary{quantikz}
\usepackage[export]{adjustbox}
\usepackage{subfig}
\usepackage[flushleft]{threeparttable}
\usepackage[ruled,vlined,linesnumbered,boxed]{algorithm2e}
\usepackage{xcolor}

\usepackage[bookmarks=true,breaklinks=true,colorlinks,citecolor=blue,linkcolor=blue,urlcolor=blue]{hyperref}

\newcommand{\CRr}{CR_r}
\newcommand{\CRb}{CR_b}

\providecommand{\BIBentrySTDinterwordspacing}{\spaceskip=0pt\relax}
\providecommand{\BIBentryALTinterwordstretchfactor}{4}
\providecommand{\BIBentryALTinterwordspacing}{\spaceskip=\fontdimen2\font plus
\BIBentryALTinterwordstretchfactor\fontdimen3\font minus
  \fontdimen4\font\relax}

\def\BibTeX{{\rm B\kern-.05em{\sc i\kern-.025em b}\kern-.08em
    T\kern-.1667em\lower.7ex\hbox{E}\kern-.125emX}}
\usepackage{authblk}
\newcommand{\keyword}[1]{Keywords: \textit{#1}}
\begin{document}

\title{QContext: Context-Aware Decomposition for Quantum Gates
}

\author[1]{Ji Liu\thanks{Corresponding Author: \texttt{ji.liu@anl.gov}}}
\author[2]{Max Bowman}
\author[3]{Pranav Gokhale}
\author[4]{Siddharth Dangwal}
\author[1]{Jeffrey Larson}
\author[3,4]{Frederic T. Chong}
\author[1]{Paul D. Hovland}

\affil[1]{Mathematics and Computer Science Division, Argonne National Laboratory}
\affil[2]{Department of Electrical \& Computer Engineering, Rice University}
\affil[3]{Super.tech, a division of ColdQuanta}
\affil[4]{Department of Computer Science, University of Chicago}

\maketitle

\begin{abstract}
In this paper we propose QContext, a new compiler structure that incorporates context-aware and topology-aware decompositions. Because of circuit equivalence rules and resynthesis, variants of a gate-decomposition template may exist. QContext exploits the circuit information and the hardware topology to select the gate variant that increases circuit optimization opportunities. We study the basis-gate-level context-aware decomposition for Toffoli gates and the native-gate-level context-aware decomposition for CNOT gates. Our experiments show that QContext reduces the number of gates as compared with the state-of-the-art approach, Orchestrated Trios~\cite{duckering2021orchestratedtrios}.
\end{abstract}

\keyword{quantum computing, circuit synthesis, compiler optimization}

\section{Introduction}
Quantum logic gates are the backbones of quantum information processing (QIP). The quantum logic gates in a program typically need to be decomposed to the basis gate set (ISA) in an assembly language. Since the target hardware may have limited connectivity, several recent works~\cite{davis2020topologyaware,younis2021qfast_topologyaware,duckering2021orchestratedtrios, niemann2022template_topologyaware} propose topology-aware decompositions that minimize the gate cost when mapping to a target hardware topology. However, these decomposition approaches use the same template to decompose the same type of quantum gates. Decomposing the gates with fixed templates lacks the opportunity to fully explore the circuit optimizations.

\begin{figure}[ht]
  \centering
    \subfloat[Canonical fully connected Toffoli decomposition resulting in ten CNOT gates\label{subfig:intro_toffoli_unopt}]{   
   \begin{adjustbox}{width=0.8\linewidth}
    \begin{quantikz}
    \qw &\ctrl{1} & \ctrl{2} & \qw &\qw &&&\ctrl{1}&\qw&\qw & \ctrl{2} & \qw &\qw &\qw &\ctrl{2}& \qw & \ctrl{1}&\gate{T} & \ctrl{1} & \qw & \qw\\
    \qw &\targ{} & \ctrl{1} &\swap{1} & \qw &=&&\targ{}&\ctrl{1} & \qw &\qw &\qw &\ctrl{1}&\qw & \qw&\gate{T}&\targ{} &\gate{T^{\dagger}} & \targ{} & \swap{1}&\qw\\
    \qw &\qw & \targ{} &\targX{}& \qw&&&\gate{H}& \targ{} &\gate{T^{\dagger}} & \targ{} & \gate{T} &\targ{} &\gate{T^{\dagger}} &\targ{}& \gate{T}& \gate{H}&\qw &\qw &\targX{}&\qw
    \end{quantikz}
    \end{adjustbox}
    }\hfill
    
    \subfloat[Context-aware Toffoli decomposition resulting in seven CNOT gates\label{subfig:intro_toffoli_opt}]{ 
    \begin{adjustbox}{width=0.8\linewidth}
    \begin{quantikz}
    \qw &\ctrl{1} & \ctrl{2} & \qw &\qw &&& \qw &\ctrl{1}\gategroup[2,steps=2,style={dashed,rounded corners, inner xsep=0.2pt},background]{{Gate Cancellation}} & \ctrl{1} & \gate{T^{\dagger}} & \ctrl{1} & \qw &\ctrl{2} &\qw & \qw &\qw& \ctrl{2}&\qw & \qw & \qw & \qw & \qw\\
    \qw &\targ{} & \ctrl{1} &\swap{1} & \qw &=&&\qw & \targ{} &\targ{} & \gate{T} &\targ{} &\gate{T^{\dagger}} &\qw &\qw & \ctrl{1} &\qw &\qw &\qw\gategroup[2,steps=4,style={dashed,rounded corners, inner xsep=0.2pt},background]{{Resynthesis to 3 CNOTs}} & \ctrl{1} & \qw &\swap{1} &\qw \\
    \qw &\qw & \targ{} &\targX{}& \qw&&&\qw & \qw &\qw &\qw &\gate{H} & \gate{T^{\dagger}} &\targ{} &\gate{T} &\targ{}& \gate{T^{\dagger}}& \targ{} &\gate{T} &\targ{} &\gate{H} & \targX{}&\qw
    \end{quantikz}
    \end{adjustbox}
    }
\caption{Two Toffoli decompositions with different total CNOT gate counts}
\vspace{-5mm}
\label{fig:intro_toffoli_context}
\end{figure}
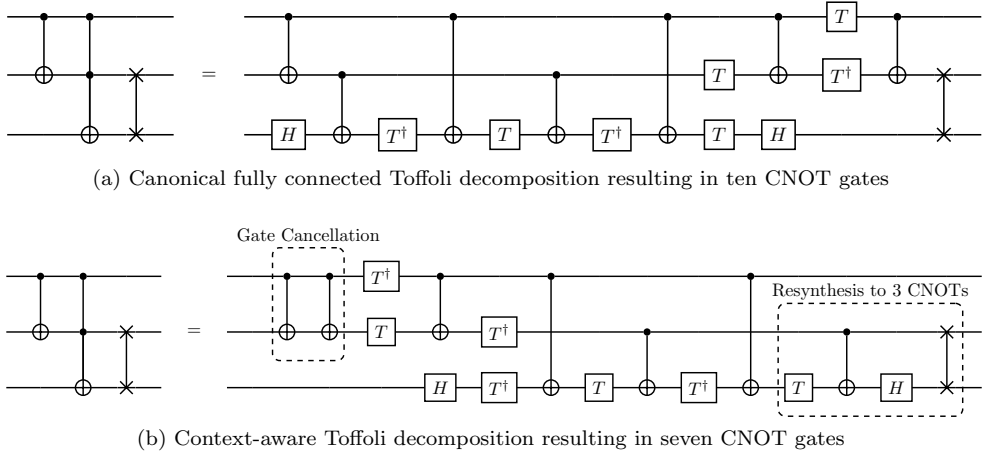

Gate context represents the predecessors and the successors in the directed acyclic graph (DAG) representation of the circuit. Since the different gate contexts could induce different circuit optimization opportunities, the quantum gate decomposition should be aware of the context. Here we use an example to demonstrate the effectiveness of context-aware gate decomposition. When the Toffoli gate in Figure~\ref{subfig:intro_toffoli_opt} is decomposed in the canonical template with six CNOT gates, there is no gate cancellation and circuit resynthesis opportunity. However, since the Toffoli gate is a self-inverse gate, inverting all gates in the canonical template will result in another decomposition template. As shown in Figure~\ref{subfig:intro_toffoli_opt}, we can perform gate cancellation and two-qubit block resynthesis optimizations when decomposing the Toffoli with the inversed template. Based on the circuit equivalence rules and resynthesis,  several variants of a gate-decomposition template might exist. %
We develop our context-aware decomposition approach QContext to identify the gate context structures and explore the circuit optimization opportunities.

QContext provides both topology-aware and context-aware decomposition. QContext first finds the gate decomposition variants compatible with the target hardware topology. Then QContext selects the best gate decomposition variant based on successors' and predecessors' information in the DAG representation of the circuit. QContext also performs context-aware decomposition when decomposing the CNOT gates to the native cross-resonance (CR) gates and single-qubit gates. We propose new templates for the CNOT gate to CR gate decomposition. We implemented QContext in Qiskit-Terra and compared it with the state-of-the-art approach Orchestrated Trios. Our experiments show that by combining the basis-gate-level and the native-level context-aware decomposition, QContext reduces both two-qubit and single-qubit gate counts in quantum circuits.

\section{Background}

\textbf{Basis Gates and Native Gates:} In order to run a quantum program on real hardware, the complex quantum gates need to be decomposed into the basis gates in an assembly language such as OpenQASM~\cite{cross2021openqasm3}. The basis gates in OpenQASM include single-qubit rotations and  two-qubit CNOT gates.

However, the CNOT gates may not be natively supported by the quantum devices. The basis gates in the assembly language should be decomposed into a sequence of native gates supported by the target hardware technology. In this paper we focus on the CR gate since it is the native gate for operating fixed-frequency superconducting devices~\cite{paraoanu2006crgate} such as IBM's superconducting systems. We expect context-aware decompositions to be applicable to other native gates such as CPHASE and iSWAP on Google~\cite{arute2019quantumsupremacy} and Rigetti~\cite{caldwell2018rigetti} devices.

\textbf{Quantum Gate Decomposition:}
Optimal quantum gate decomposition is crucial for quantum compilation. Extensive research~\cite{vartiainen2004synthesis,mottonen2004synthesis2, shende2006synthesis3, davis2020topologyaware} has been done to find efficient decompositions for unitary matrices.
Besides the generalized decomposition, researchers have proposed structured templates~\cite{jones2013lowToffoliTgate, cleve2000parallelQFT} to decompose certain types of quantum gates. 

With the recent development in quantum hardware, several researchers have  proposed topology-aware decompositions~\cite{hu2019mctsquare, cheng2018toffolilinear} that minimize the gate cost when mapping to a target hardware topology. Toffoli gate is one of the most commonly used three-qubit gates in quantum information processing. Duckering et al.~\cite{duckering2021orchestratedtrios} proposed a compiler structure Orchestrated Trios, which efficiently routes and decomposes the Toffoli gates. They observed that it is more efficient to preserve the Toffoli gate during routing instead of routing each individual CNOT gate in the Toffoli gate decomposition. The approach first decomposes the quantum program to the one-, two-, and three-qubit gates and routes the gates. %
Then a second decomposition step is performed, which decomposes the Toffoli gate according to the connectivity of the physical qubits. %

\textbf{Quantum Optimization:}
The state-of-the-art quantum compilers mostly exploit circuit optimizations at the basis gate level. The Qiskit~\cite{Qiskit}, $t\ket{ket}$~\cite{sivarajah2020tket}, and Cirq~\cite{cirq} compilers contain optimization passes that identify multiqubit blocks and resynthesize the blocks with KAK decomposition~\cite{kraus2001kak1} to reduce circuit cost. Gate cancellation~\cite{maslov2008gatecancellation} is another useful optimization. Quantum gates may commute, and the compiler can reorder the gates to optimize the circuits and improve routing with gate cancellation~\cite{liu2022NASSC}. %
Other optimizations include cross-talk mitigation~\cite{murali2020softwarecrosstalk}, peephole optimization~\cite{rpo}, and dynamical decoupling~\cite{smith2022timestitch}.

\section{QContext}
\label{sec:qcontext}
\begin{figure}[htbp]
\centering
    \includegraphics[width = 0.5\linewidth]{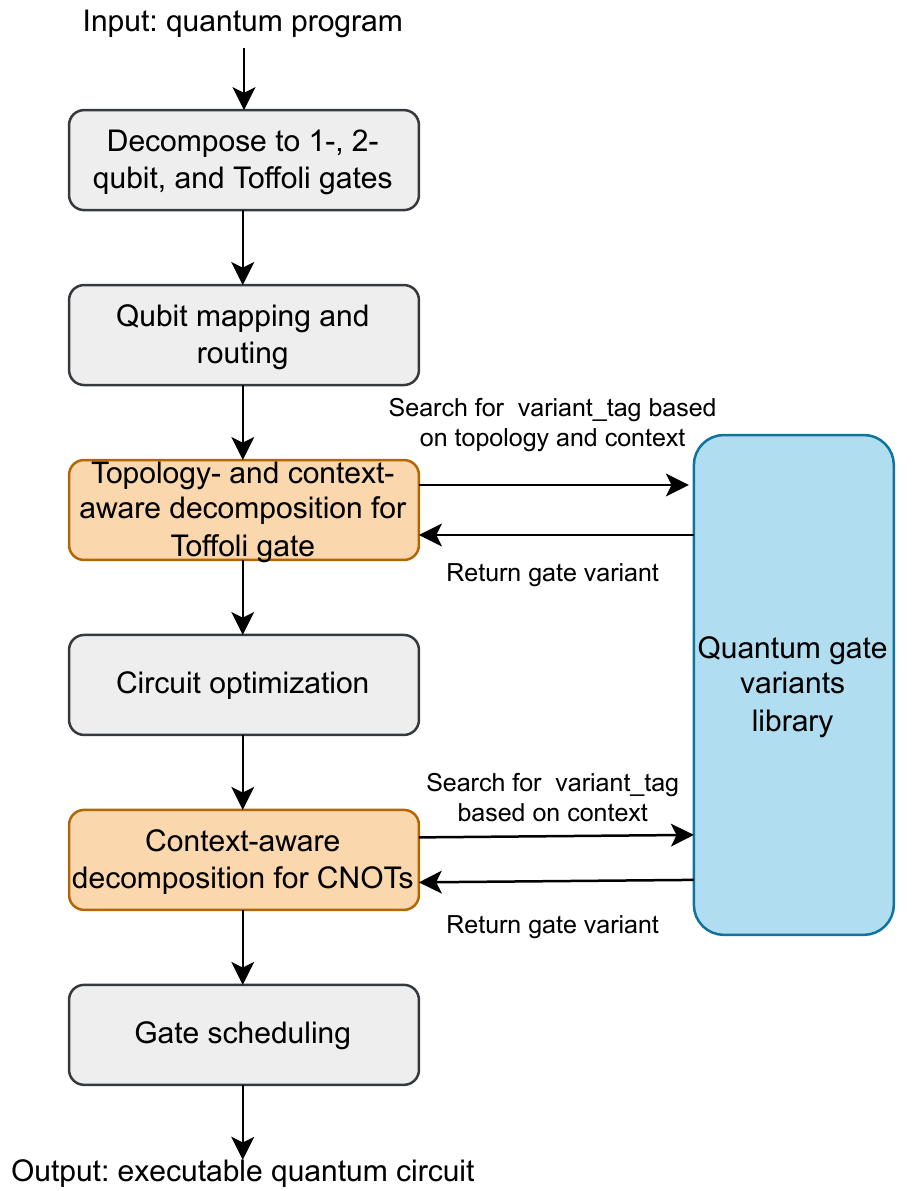}
  \caption{Compilation flow of QContext. The steps introduced by QContext are marked with an orange background.}
\label{fig:compileflow}
\end{figure}

QContext performs basis-gate-level context-aware decomposition for the Toffoli gates and performs  native-gate-level context-aware decomposition for the CNOT gates. Figure~\ref{fig:compileflow} shows the compilation flow of QContext. The boxes with the grey color backgrounds are the original compilation steps in Trios and Qiskit. The boxes with the orange color backgrounds are the compilation steps introduced and modified in QContext. Given an input quantum program, QContext first decomposes the gates to 1-qubit gates, 2-qubit gates, and Toffoli gates. Not decomposing the Toffoli gate allows the routing algorithm to capture the program structure and reduce the routing overhead~\cite{duckering2021orchestratedtrios}. The compiler performs qubit mapping and routing for the gates. We propose a gate library that contains the gate decomposition variants for the Toffoli and CNOT gates. The gate library contains 32 Toffoli gate variants and 14 CNOT gate variants. Each gate variant is associated with a variant$\_$tag. The compiler specifies searches in the library and returns the best matching gate variant based on the hardware topology and the gate context. Then, the compiler performs circuit optimizations. After the optimizations, the basis gates need to be decomposed into native gates. For each CNOT gate in the circuit, the compiler finds the CNOT gate variant that minimizes the number of single-qubit rotations after optimization.

\section{Basis-gate-level decomposition}
We use the Toffoli gate as an example to discuss the format and the strategies for generating gate variants.
\label{sec:basis-gate-level}
\subsection{Variant$\_$tag}
We label the gate variants with different tags to differentiate them. The variant$\_$tag for an n-qubit gate is a tuple with five elements:
\begin{equation} 
\textit{variant\_tag} = \textit{(pre\_tag, suc\_tag, design\_tag, topo\_tag, opt\_tag)}
\end{equation}

The first element \textit{pre$\_$tag} indicates the position of the CNOT gates at the beginning of the circuit that can be canceled or optimized with the predecessors. As shown in Figure~\ref{fig:toffoli_permute}, the first CNOT gate is between q0 and q2. The \textit{pre$\_$tag} is set to 
``02'' for this gate variant. The second element \textit{suc$\_$tag} indicates the position of the CNOTs at the end of the circuit that can be canceled or optimized with the successors. The last two CNOTs in the figure are between q0 and q1, and the \textit{suc$\_$tag} is set to ``10.'' The order of the bits indicates the control qubit and the target qubit.

The \textit{design$\_$tag} is an integer that indicates the basic design of the template. The gate variants that are generated by applying the equivalence rules to the canonical Toffoli decomposition will have the same \textit{design$\_$tag = 0}. However, the gate variants based on resynthesis will have different designs and should have different \textit{design$\_$tag}. The \textit{topo$\_$tag} suggests the connectivity of the physical qubits. \textit{topo$\_$tag} can have four values: F, L0, L1, and L2. ``F'' or ``L'' indicates that the three qubits are fully connected or linearly connected. The integer represents the id of the qubit that is connected to the other two qubits. The last element \textit{opt$\_$tag} is used to differentiate the gate variants with similar structures but with different optimization opportunities. \textit{opt$\_$tag} can be either ``O'' or ``I,'' which differentiates the original decomposition and the inversed decomposition. The variant$\_$tag provides an efficient way of defining and searching gate variants. 

\begin{figure}[htbp]
  \centering
    \begin{adjustbox}{width=0.8\linewidth}
    \begin{quantikz}
    \lstick{$q_0$} &\qw &\ctrl{2} &\qw & \qw & \qw &\ctrl{2} &\qw & \qw &\gate{T}& \targ{}&\gate{T^{\dagger}} & \targ{} & \qw\\
    \lstick{$q_1$} &\qw &\qw & \qw &\ctrl{1}&\qw &\qw&\qw & \ctrl{1}&\qw&\ctrl{-1} &\gate{T} & \ctrl{-1} & \qw\\
    \lstick{$q_2$} &\gate{H}& \targ{} &\gate{T^{\dagger}} & \targ{} & \gate{T} &\targ{} &\gate{T^{\dagger}} &\targ{}& \gate{T}& \gate{H}&\qw &\qw &\qw
    \end{quantikz}
    \end{adjustbox}
\captionsetup{font={small}}
\caption{Fully connected Toffoli variant by permuting control qubits, variant$\_$tag = $(02, 10, 0,F,O)$}
\label{fig:toffoli_permute}
\end{figure}

\subsection{Toffoli Gate Variants}
\label{subsec:equivalence}
We can generate the gate variants based on the circuit equivalence rules~\cite{garcia2011equivalent} and resynthesis. %
We used three strategies described below to find gate variants.

\textbf{1) Self-inverse:} For the gates that are self-inverse, we can place the gates in the decomposition template in a reversed order and inverse each gate to get a new gate variant. Since the front layer of the circuit permutes with the last layer of the circuit, the variant\_tag after inversion is specified by permuting $pre\_tag$ with $suc\_tag$. %
The \textit{opt$\_$tag} changes from ``O'' to ``I.'' Other self-inverse gates include the CNOT gate, SWAP gate, Bridge gate~\cite{itoko2020bridge}, and Fredkin gate~\cite{brown2001fredkin}.

\textbf{2) Permute control and target qubits:} For a gate with multiple control qubits, we can permute the control qubits in a decomposition template to generate new gate variants. Figure~\ref{fig:toffoli_permute} shows the gate variant that is generated based on the canonical decomposition and the permutation of control qubits $q0, q1$. The $variant\_tag$ is specified by switching the associated qubits in the $pre\_tag$ and $suc\_tag$. The $design\_tag$, $topo\_tag$, and $opt\_tag$ remain to be the same after permutation.

When the target gate is a NOT gate, we can also permute the control qubit with the target qubit by converting the multicontrolled NOT gate to the multicontrolled Z gate~\cite{garcia2011equivalent}. %
The multicontrolled Z gates are symmetrical; thus any qubit can be the target qubit.

\textbf{3) Resynthesis:}
Based on equivalence rules, we can find many gate variants. However, some structures  cannot be generated with the equivalence rules. One solution is to leverage the gate decomposition templates in prior works. Another solution is to resynthesize~\cite{davis2020topologyaware, younis2021qfast_topologyaware,smith2021leap} the unitary matrix to the quantum circuit with the expected structures. A new $design\_tag$ will be assigned to the synthesized template.

\begin{figure}[htbp]
  \centering
   \begin{adjustbox}{width=0.8\linewidth}
    \begin{quantikz}
    \lstick{$q_0$} &\qw &\qw & \ctrl{2} &\qw & \qw & \qw & \ctrl{2} &\qw & \ctrl{2} & \qw & \qw & \ctrl{2} & \gate{T^{\dagger}} & \qw & \qw \\
    \lstick{$q_1$} &\qw & \targ{} & \qw & \gate{T} & \targ{} & \gate{T} & \qw & \targ{} & \qw & \gate{T^{\dagger}} & \targ{} & \qw & \gate{T^{\dagger}} & \qw & \qw \\
    \lstick{$q_2$} &\gate{H} & \ctrl{-1} & \targ{} & \qw & \ctrl{-1} & \gate{T} & \targ{} & \ctrl{-1} & \targ{} & \qw & \ctrl{-1} & \targ{} & \gate{T^{\dagger}} &\gate{H} &\qw
    \end{quantikz}
    \end{adjustbox}
\captionsetup{font={small}}
\caption{Toffoli gate variant with linear connectivity, variant$\_$tag = $(21, 02, 1, L2, O )$}
\label{fig:Toffoli_linear}
\end{figure}
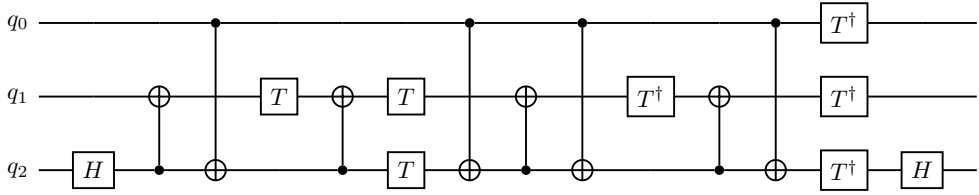

We can apply the equivalence rules to the canonical 8-CNOT linear Toffoli decomposition to generate gate variants with linear connectivity. The 8-CNOT linear Toffoli decomposition is assigned with a $design\_tag$ equal to 1, since it is different from the fully connected design. 
Figure~\ref{fig:Toffoli_linear} shows one of the linear Toffoli gate variants. Since q2 is connected to q0 and q1, the \textit{topo$\_$tag} is set to ``L2.'' 

\subsection{Variant$\_$tag Calculation}
Here we describe the method to specify the variant$\_$tag based on hardware topology and gate context. %
For each Toffoli gate node in the DAG, the compiler first determines the \textit{topo$\_$tag} based on the connectivity of the physical qubits. Then for each gate variant in the library with the \textit{topo$\_$tag}, we can estimate the CNOT gate count reduction after optimization. Since the number of three-qubit gate variants is limited, the compiler performs an exhaustive search in the library and returns the gate variant with the maximum CNOT gate count reduction.

\section{Native-gate-level decomposition} 
\label{sec:native-gate-level}

\subsection{Variant$\_$tag}
\begin{figure}[ht]
  \centering
   \begin{adjustbox}{width=0.6\linewidth}
    \begin{quantikz}
    \qw &\ctrl{1} &\qw &\Rightarrow &&\gate{U3(\theta_1, \phi_1, \lambda_1)}&\gate[wires=2]{\uparrow\downarrow CR_{+-}}& \gate{U3(\theta_2, \phi_2, \lambda_2)}&\qw\\
    \qw &\targ{} &\qw && &\gate{U3(\theta_3, \phi_3, \lambda_3)} & & \gate{U3(\theta_4, \phi_4, \lambda_4)}&\qw 
    \end{quantikz} 
    \end{adjustbox}
\caption{Generalized CNOT gate decomposition template}
\label{fig:CNOT_general}
\end{figure}
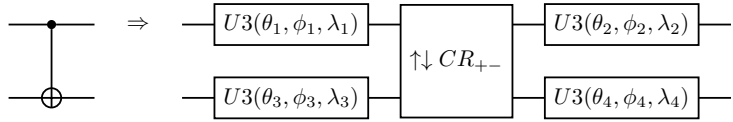

First, we introduce the format of the CNOT gate variant$\_$tag. Two  orientations of the CR gate are possible. The generalized CNOT gate decomposition circuit is shown in Figure~\ref{fig:CNOT_general}. The U3 gate is the generic single-qubit rotation gate with three angles $\theta,\phi$, and $\lambda$. Since the physical qubit connectivity for the CNOT gate is always linear, we do not need the \textit{topo\_tag} and \textit{opt\_tag}. The CNOT gate variant tag is a tuple with twelve elements that specify the four U3 gates in the template and an \textit{ori$\_$tag} to specify the orientation of the CR gate.
\begin{equation} 
\begin{aligned}
\textit{variant\_tag} = (\theta_1,\phi_1,\lambda_1, ... ,\theta_4,\phi_4,\lambda_4, ori\_tag)\nonumber
\end{aligned}
\end{equation}
The variant$\_$tag allows us to quickly find the inverse gate and calculate the native gate count. $U3(-\theta,-\lambda,-\phi)$ is the inverse gate of $U3(\theta,\phi,\lambda)$. The generalized U3 gate requires two $Rx(90)$ pulses and three Rz gates. The Rz gates~\cite{mckay2017virtualz} are implemented in software by frame change and do not introduce any noise. When decomposing the CNOT gate, the compiler first specifies the \textit{ori$\_$tag} based on the orientation of the target physical qubit connection. Then, for each variant$\_$tag that contains the correct \textit{ori$\_$tag}, the compiler estimates the total number of $Rx(90)$ pulses after optimization and selects the one with the fewest $Rx(90)$ gates.

\subsection{CNOT Gate Variants}

We have three strategies for finding CNOT gate variants.
\textbf{1) Self-inverse:} The CNOT gate is a self-inverse gate. We can place the native gates in a reversed order and invert each native gate to get a new gate decomposition variant. %

\textbf{2) Polarity switch:} The CR native gate is implemented with a positive half-CR and a negative half-CR gate to mitigate noise~\cite{sundaresan2020echocr}. We can instead implement the CR gate as a negative half-CR followed by a positive half-CR gate. This polarity switch~\cite{gokhale2021swap, bowman2022hardware} introduces a side effect that appends two $Rx(180)$ gates to the left and the right of the CR native gate. We can also combine the polarity switch and the gate inverse. %

\textbf{3) Resynthesis:} We used a numerical optimization approach to find novel CNOT gate decompositions. We may treat the discovery of novel decomposition of the CNOT unitary matrix as a nonlinear least-squares optimization problem. An objective function can be defined as the square of the Frobenius norm of the difference between the parameter-generated unitary matrix and the target unitary matrix, where the U3 parameters are adjusted to minimize the objective function. Several standard least-squares optimization algorithms such as Levenberg--Marquardt\cite{more1978levenberg} and Broyden--Fletcher--Goldfarb--Shanno~\cite{head1985bfgs} can be used. %

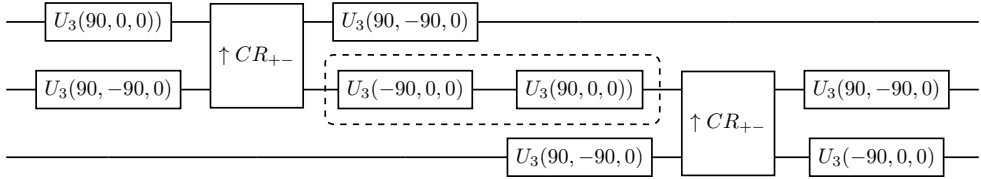
\begin{figure}[htbp]
        \centering
   \begin{adjustbox}{width=0.8\linewidth}
    \begin{quantikz}
    \qw & \gate{U_3(90,0,0))}&\gate[wires=2]{\uparrow CR_{+-}} &\gate{U_3(90,-90,0)}&\qw &\qw & \qw & \qw\\
    \qw & \gate{U_3(90,-90,0)} & & \gate{U_3(-90,0,0)}\gategroup[1,steps=2,style={dashed,rounded corners, inner xsep=0.2pt},background]{{}}& \gate{U_3(90,0,0))}&\gate[wires=2]{\uparrow CR_{+-}} &\gate{U_3(90,-90,0)} & \qw \\
    \qw & \qw & \qw &\qw& \gate{U_3(90,-90,0)} & & \gate{U_3(-90,0,0)} & \qw 
    \end{quantikz} 
    \end{adjustbox}
    
     \caption{Ladder-shaped CNOT structure and gate decomposition}
\label{fig:ladder}
\end{figure}

We found gate variants that have a similar number of $Rx(90)$ rotations as the canonical decomposition but with more gate cancellation opportunities. For instance, the ladder-shaped circuit structure shown in Figure~\ref{fig:ladder} widely exists in quantum circuits~\cite{kandala2017hardwareefficient,itoko2020bridge,li2022paulihedral}. Here we use a numerical approach to find the gate variant that has $U_3(90,0,0)$ on the top left and the inverse gate $U_3(-90,0,0)$ on the bottom right corner. When there is a sequence of ladder-shaped CNOTs, the single-qubit gates in between will cancel out. In Figure~\ref{fig:ladder}, the single-qubit gates that will cancel out are marked with a dashed box. Based on the numerical approach, we obtained six CNOT gate variants that have the self-cancellation property.  

\section{Methodology}
\label{sec:methodology}

\textbf{Implementation:} We implement our QContext framework on the open-source quantum computing framework Qiskit~\cite{Qiskit}. The version of Qiskit-Terra is 0.18.3, and our implementation is publicly available at \url{https://github.com/revilooliver/context-aware-decomposition}. We compare our results with the Orchestrated Trios compiler implemented in Qiskit.

\textbf{Benchmarks:} The benchmarks in our experiments are derived from the Toffoli benchmarks in Orchestrated Trios~\cite{triosbenchmark} and the reversible circuits~\cite{revlib}. Besides the benchmarks with Toffoli gates, we  include the VQE UCCSD ansatz for LiH~\cite{peruzzo2014uccsd}  and QAOA for Max-Cut~\cite{farhi2014qaoa}. We use these two benchmarks to study the performance of CNOT gate decomposition. 

\section{Evaluation and Discussion}
\label{sec:evaluation}
In this section we show the single- and two-qubit gate numbers using QContext compared with Trios on \texttt{ibmq\_montreal} which has heavy-hex topology~\cite{takitaibmheavyhex}. 

The CR gate and single-qubit gate reductions are shown in Table~\ref{table:result_montreal}. All the Toffoli gates are decomposed into the linearly connected gate variants with 8 CNOTs. Since our optimization focuses on the Toffoli gates and the CNOT gates, to distinguish the effect of qubit routing and our approach, we separate the CR gates for qubit routing and the CR gates for the benchmark. $\CRr$ represents the CR gates that form the SWAP gate and are not involved in the optimization. A better routing algorithm may reduce the $\CRr$ number, but that is beyond the scope of the current discussion. %
$\CRb$ represents the number of CR gates that are needed for implementing the benchmark. $\Delta \CRb$ is the percentage change in the number of $\CRb$: $\Delta \CRb$ = 1-$\CRb(QContext)/\CRb(Trios)$. $\Delta \text{\#sx}$ is the percentage change in the number of total sx rotations: $\Delta \text{\#sx}= 1 - \text{\#sx}(\text{QContext})/\text{\#sx}(\text{Trios})$. In the native gates, the z-axis rotation Rz gates are implemented in software by frame change; only the 90-degree x-axis rotation $sx$ and the 180-degree x-axis rotation $x$ gate will introduce noise. Here we calculate the total number of x-axis 90-degree rotations  as $\#sx$. An $x$ gate is counted as two $sx$ gates. The CNOT gate context-aware decomposition also optimizes the inserted SWAP gates, so we do not differentiate the routing circuit when calculating $\#sx$. 

\begin{table}[bhtp]
  \caption{Number of CR gates of QContext in comparison with Trios~\cite{duckering2021orchestratedtrios} on \texttt{ibmq\_montreal}}
  \label{table:result_montreal}
  \centering
  \small
    \begin{threeparttable}
  \begin{tabular}{|c|c|c|c|c|c|c|c|c|c|}
  \hline
  \multicolumn{3}{|c|}{\textbf{Circuit}} &  \multicolumn{2}{c|}{\textbf{Trios}} &  \multicolumn{2}{c|}{\textbf{QContext}} &  \multicolumn{3}{c|}{\textbf{Comparison}}\\
    \hline
     \textbf{benchmark} & $\#Q$ & $\CRr$ & $\CRb$ & $\#sx$ &$\CRb$ & $\#sx$ & $\Delta \CRb$ & $\Delta sx$ & $T_c/T_t$\\
    \hline
cnx\_half~\cite{cnx-halfborrowed} & 5 & 0 & 35 & 112 & 29 & 64 & 17.1\% & 42.9\% & 1.14\\\hline
sym6~\cite{revlib} & 10 & 99 & 128 & 631 & 100 & 402 & 21.9\% & 21.7\% & 1.09\\\hline
cnx\_dirty~\cite{baker2019cnxdirty} & 11 & 102 & 146 & 666 & 119 & 520 & 18.5\% & 21.9\% & 1.02\\\hline
cnx\_half~\cite{cnx-halfborrowed} & 19 & 132 & 289 & 1204 & 230 & 841 & 20.4\% & 30.1\% & 1.26\\\hline
cnx\_log~\cite{barenco1995cnx-logancilla} & 19 & 195 & 163 & 972 & 129 & 799  & 20.9\% & 17.8\% & 1.32 \\\hline
c\_adder~\cite{cuccaro2004adder} & 20 & 507 & 196 & 1828 & 184 & 1579 & 6.1\% & 13.6\% & 1.19\\\hline
t\_adder~\cite{takahashi2009adder} & 20 & 483 & 220 & 1859 & 191 & 1532 & 13.2\% & 17.6\% & 1.04 \\\hline
incrementer~\cite{cnx-halfborrowed} & 5 & 105 & 461 & 1585 & 421 & 1114 &  8.7\% & 29.7\%  & 1.02 \\\hline
grover~\cite{grover1996grover} & 9 & 201 & 724 & 2524 & 609 & 1848  & 15.9\%  &26.8\%  & 1.16 \\\hline
QAOA~\cite{farhi2014qaoa} & 10 & 168 & 96 & 672 & 96 & 592  & --- & 11.9\%  & 1.04 \\\hline
UCCSD~\cite{peruzzo2014uccsd} & 8 & 654 & 274 & 2348& 274 & 2022 & ---  &13.9\%  & 1.10 \\\hline

\multicolumn{1}{|c|}{\textbf{Geo mean}} & \multicolumn{6}{c|}{} & \multicolumn{1}{c|}{14.7\%} & \multicolumn{1}{c|}{20.9\%} & \multicolumn{1}{c|}{1.12} \\ \hline

  \end{tabular}
      \begin{tablenotes}
      \item \#Q denotes the number of qubits. t\_adder is the takahashi\_adder~\cite{takahashi2009adder}. c\_adder is the cuccaro\_adder~\cite{cuccaro2004adder}.
      \item $T_c/T_t$ is the ratio between total transpilation time of QContext and Trios. %
    \end{tablenotes}
  \end{threeparttable}
\end{table}

As shown in Table~\ref{table:result_montreal}, QContext reduces the number of single-qubit rotations $\#sx$ for all the benchmarks and reduces $\CRb$ for all the benchmarks that contain Toffoli gates. The geometric mean of $\Delta \CRb$ for the benchmarks that contain Toffoli gates is $14.7\%$. The geometric mean of $\Delta sx$ is $20.9\%$. The results for the QAOA benchmark and the UCCSD benchmark show the effectiveness of CNOT gate context-aware decomposition. Note that QAOA and the UCCSD benchmark do not contain any Toffoli gate, therefore, they will not have any CNOT gate reduction. We also compare the compilation time. The compilation time is the average of ten runs. The choice of using a library of pre-synthesized templates makes QContext more scalable in terms of compilation time. %
The average compilation time on \texttt{ibmq$\_$montreal} increases by $12\%$ compared with that of Trios. 

\section{Conclusion}
\label{sec:conclusion}
In this paper we propose QContext, a new compiler structure that incorporates context-aware and topology-aware decompositions. We highlight that quantum gate decomposition should be aware of the context. QContext reduces both the single- and two-qubit gate counts by being aware of the basis-gate-level and native-gate-level gate context. 

\section{Acknowledgement} \label{sec:acknowledgement}
This material is based upon work supported by Q-NEXT, one of the U.S. Department of Energy Office of Science (DOE-SC) National Quantum Information Science Research Centers and the Office of Advanced Scientific Computing Research, Accelerated Research for Quantum Computing program. This work is funded in part by EPiQC, an NSF Expedition
in Computing, under award CCF-1730449; in part
by STAQ under award NSF Phy-1818914; 
in part by the US Department of Energy Office 
of Advanced Scientific Computing Research, Accelerated 
Research for Quantum Computing Program; and in part by the 
NSF Quantum Leap Challenge Institute for Hybrid Quantum Architectures and 
Networks (NSF Award 2016136) and in part based upon work supported by the 
U.S. Department of Energy, Office of Science, National Quantum 
Information Science Research Centers. This research used resources of the Oak Ridge Leadership Computing Facility, which is a DOE Office of Science User Facility supported under Contract DE-AC05-00OR22725.
FTC is Chief Scientist for Quantum Software at Infleqtion and an advisor to Quantum Circuits, Inc.

\vfill
\framebox{\parbox{.90\linewidth}{\scriptsize The submitted manuscript has been
created by UChicago Argonne, LLC, Operator of Argonne National Laboratory
(``Argonne''). Argonne, a U.S.\ Department of Energy Office of Science
laboratory, is operated under Contract No.\ DE-AC02-06CH11357.  The U.S.\
Government retains for itself, and others acting on its behalf, a paid-up
nonexclusive, irrevocable worldwide license in said article to reproduce,
prepare derivative works, distribute copies to the public, and perform publicly
and display publicly, by or on behalf of the Government.  The Department of
Energy will provide public access to these results of federally sponsored
research in accordance with the DOE Public Access Plan
\url{http://energy.gov/downloads/doe-public-access-plan}.}}


\begin{thebibliography}{00}

\bibitem{Qiskit}ANIS, M., et al., Qiskit: An Open-source Framework for Quantum Computing.  (2021)

\bibitem{arute2019quantumsupremacy}
F.~Arute, et al., ``Quantum supremacy using a
  programmable superconducting processor,'' \emph{Nature}, vol. 574, no. 7779,
  pp. 505--510, 2019.

\bibitem{baker2019cnxdirty}
J.~M. Baker, C.~Duckering, A.~Hoover, and F.~T. Chong, ``Decomposing quantum
  generalized {T}offoli with an arbitrary number of ancilla,''
  \emph{arXiv:1904.01671}, 2019.

\bibitem{barenco1995cnx-logancilla}
A.~Barenco, C.~H. Bennett, R.~Cleve, D.~P. DiVincenzo, N.~Margolus, P.~Shor,
  T.~Sleator, J.~A. Smolin, and H.~Weinfurter, ``Elementary gates for quantum
  computation,'' \emph{Physical Review A}, vol.~52, no.~5, p. 3457, 1995.

\bibitem{brown2001fredkin}
J.~Brown, \emph{Quest for the quantum computer}.\hskip 1em plus 0.5em minus
  0.4em\relax Simon and Schuster, 2001.


\bibitem{caldwell2018rigetti}
S.~A. Caldwell, et al., ``Parametrically activated
  entangling gates using transmon qubits,'' \emph{Physical Review Applied},
  vol.~10, no.~3, p. 034050, 2018.

\bibitem{cheng2018toffolilinear}
X.~Cheng, Z.~Guan, and W.~Ding, ``Mapping from multiple-control {Toffoli}
  circuits to linear nearest neighbor quantum circuits,'' \emph{Quantum
  Information Processing}, vol.~17, no.~7, pp. 1--26, 2018.

\bibitem{cirq}
\BIBentryALTinterwordspacing
{Cirq developers}, ``Cirq,'' 2021. [Online]. Available:
  \url{https://zenodo.org/record/5182845}
\BIBentrySTDinterwordspacing

\bibitem{cross2021openqasm3}
A.~W. Cross, A.~Javadi-Abhari, T.~Alexander, N.~de~Beaudrap, L.~S. Bishop,
  S.~Heidel, C.~A. Ryan, J.~Smolin, J.~M. Gambetta, and B.~R. Johnson,
  ``{OpenQASM} 3: A broader and deeper quantum assembly language,''
  \emph{arXiv:2104.14722}, 2021.

\bibitem{cuccaro2004adder}
S.~A. Cuccaro, T.~G. Draper, S.~A. Kutin, and D.~P. Moulton, ``A new quantum
  ripple-carry addition circuit,'' \emph{arXiv quant-ph/0410184}, 2004.


\bibitem{davis2020topologyaware}
M.~G. Davis, E.~Smith, A.~Tudor, K.~Sen, I.~Siddiqi, and C.~Iancu, ``Towards
  optimal topology aware quantum circuit synthesis,'' in \emph{International
  Conference on Quantum Computing and Engineering}.\hskip 1em plus 0.5em minus
  0.4em\relax IEEE, 2020, pp. 223--234.

\bibitem{triosbenchmark}
\BIBentryALTinterwordspacing
C.~Duckering, J.~M. Baker, P.~Gokhale, and A.~Litteken, ``Quantum circuit
  benchmarks,'' 2020. [Online]. Available:
  \url{https://github.com/jmbaker94/quantumcircuitbenchmarks}
\BIBentrySTDinterwordspacing

\bibitem{duckering2021orchestratedtrios}
C.~Duckering, J.~M. Baker, A.~Litteken, and F.~T. Chong, ``Orchestrated
  {T}rios: compiling for efficient communication in quantum programs with
  3-qubit gates,'' in \emph{Proceedings of the 26th ACM International
  Conference on Architectural Support for Programming Languages and Operating
  Systems}, 2021, pp. 375--385.

\bibitem{farhi2014qaoa}
E.~Farhi, J.~Goldstone, and S.~Gutmann, ``A quantum approximate optimization
  algorithm,'' \emph{arXiv:1411.4028}, 2014.

\bibitem{garcia2011equivalent}
J.~C. Garcia-Escartin and P.~Chamorro-Posada, ``Equivalent quantum circuits,''
  \emph{arXiv:1110.2998}, 2011.

\bibitem{cnx-halfborrowed}
\BIBentryALTinterwordspacing
C.~Gidney, ``Constructing large controlled {NOT}s,'' 2015. [Online]. Available:
  \url{https://algassert.com/circuits/2015/06/05/Constructing-Large-Controlled-Nots.html}
\BIBentrySTDinterwordspacing

\bibitem{gokhale2021swap}
P.~Gokhale, T.~Tomesh, M.~Suchara, and F.~T. Chong, ``Faster and more reliable
  quantum {SWAPs} via native gates,'' \emph{arXiv:2109.13199}, 2021.


\bibitem{bowman2022hardware}
M.~A. Bowman, P.~Gokhale, J.~Larson, J.~Liu, and M.~Suchara,
  ``Hardware-conscious optimization of the quantum Toffoli gate,'' 
  \emph{arXiv:2209.02669}, 2022.
  
\bibitem{grover1996grover}
L.~K. Grover, ``A fast quantum mechanical algorithm for database search,'' in
  \emph{Proceedings of the Twenty-Eighth Annual ACM Symposium on Theory of
  Computing}, 1996, pp. 212--219.

\bibitem{gui2020term}
K.~Gui, T.~Tomesh, P.~Gokhale, Y.~Shi, F.~T. Chong, M.~Martonosi, and
  M.~Suchara, ``Term grouping and travelling salesperson for digital quantum
  simulation,'' \emph{arXiv:2001.05983}, 2020.

\bibitem{head1985bfgs}
J.~D. Head and M.~C. Zerner, ``A {Broyden--Fletcher--Goldfarb--Shanno}
  optimization procedure for molecular geometries,'' \emph{Chemical Physics
  Letters}, vol. 122, no.~3, pp. 264--270, 1985.

\bibitem{hu2019mctsquare}
S.~Hu, D.~Maslov, M.~Pistoia, and J.~Gambetta, ``Efficient circuits for quantum
  search over {2D} square lattice architecture,'' in \emph{Proceedings of the
  56th Annual Design Automation Conference}, 2019, pp. 1--2.

\bibitem{itoko2020bridge}
T.~Itoko, R.~Raymond, T.~Imamichi, and A.~Matsuo, ``Optimization of quantum
  circuit mapping using gate transformation and commutation,''
  \emph{Integration}, vol.~70, pp. 43--50, 2020.

\bibitem{kandala2017hardwareefficient}
A.~Kandala, A.~Mezzacapo, K.~Temme, M.~Takita, M.~Brink, J.~M. Chow, and J.~M.
  Gambetta, ``Hardware-efficient variational quantum eigensolver for small
  molecules and quantum magnets,'' \emph{Nature}, vol. 549, no. 7671, pp.
  242--246, 2017.


\bibitem{kraus2001kak1}
B.~Kraus and J.~Cirac, ``Optimal creation of entanglement using a two-qubit
  gate,'' \emph{Physical Review A}, vol.~63, no.~6, p. 062309, 2001.

\bibitem{li2022paulihedral}
G.~Li, A.~Wu, Y.~Shi, A.~Javadi-Abhari, Y.~Ding, and Y.~Xie, ``Paulihedral: A
  generalized block-wise compiler optimization framework for quantum simulation
  kernels,'' in \emph{Proceedings of the 27th ACM International Conference on
  Architectural Support for Programming Languages and Operating Systems}, 2022,
  pp. 554--569.


\bibitem{liu2022NASSC}
J.~Liu, P.~Li, and H.~Zhou, ``Not all swaps have the same cost: A case for
  optimization-aware qubit routing,'' in \emph{2022 IEEE International
  Symposium on High-Performance Computer Architecture (HPCA)}.\hskip 1em plus
  0.5em minus 0.4em\relax IEEE, 2022, pp. 709--725.
  
\bibitem{maslov2008gatecancellation}
D.~Maslov, G.~W. Dueck, D.~M. Miller, and C.~Negrevergne, ``Quantum circuit
  simplification and level compaction,'' \emph{IEEE Transactions on
  Computer-Aided Design of Integrated Circuits and Systems}, vol.~27, no.~3,
  pp. 436--444, 2008.


\bibitem{mckay2017virtualz}
D.~C. McKay, C.~J. Wood, S.~Sheldon, J.~M. Chow, and J.~M. Gambetta,
  ``Efficient ${Z}$ gates for quantum computing,'' \emph{Physical Review A},
  vol.~96, no.~2, p. 022330, 2017.

\bibitem{more1978levenberg}
J.~J. Mor{\'e}, ``The {Levenberg}--{Marquardt} algorithm: Implementation and
  theory,'' in \emph{Numerical Analysis}.\hskip 1em plus 0.5em minus
  0.4em\relax Springer, 1978, pp. 105--116.

\bibitem{mottonen2004synthesis2}
M.~M{\"o}tt{\"o}nen, J.~J. Vartiainen, V.~Bergholm, and M.~M. Salomaa,
  ``Quantum circuits for general multiqubit gates,'' \emph{Physical Review
  Letters}, vol.~93, no.~13, p. 130502, 2004.


\bibitem{murali2020softwarecrosstalk}
P.~Murali, D.~C. McKay, M.~Martonosi, and A.~Javadi-Abhari, ``Software
  mitigation of crosstalk on noisy intermediate-scale quantum computers,'' in
  \emph{Proceedings of the Twenty-Fifth International Conference on
  Architectural Support for Programming Languages and Operating Systems}, 2020,
  pp. 1001--1016.

\bibitem{rpo}
J.~Liu, L.~Bello, and H.~Zhou, ``Relaxed peephole optimization: A novel
  compiler optimization for quantum circuits,'' in \emph{2021 IEEE/ACM
  International Symposium on Code Generation and Optimization (CGO)}.\hskip 1em
  plus 0.5em minus 0.4em\relax IEEE, 2021, pp. 301--314.
  
\bibitem{smith2022timestitch}
K.~N. Smith, G.~S. Ravi, P.~Murali, J.~M. Baker, N.~Earnest, A.~Javadi-Cabhari,
  and F.~T. Chong, ``Timestitch: Exploiting slack to mitigate decoherence in
  quantum circuits,'' \emph{ACM Transactions on Quantum Computing}, vol.~4,
  no.~1, pp. 1--27, 2022.


\bibitem{niemann2022template_topologyaware}
P.~Niemann, A.~A. de~Almeida, G.~Dueck, and R.~Drechsler, ``Template-based
  mapping of reversible circuits to {IBM} quantum computers,''
  \emph{Microprocessors and Microsystems}, vol.~90, p. 104487, 2022.


\bibitem{paraoanu2006crgate}
G.~Paraoanu, ``Microwave-induced coupling of superconducting qubits,''
  \emph{Physical Review B}, vol.~74, no.~14, p. 140504, 2006.

\bibitem{peruzzo2014uccsd}
A.~Peruzzo, J.~McClean, P.~Shadbolt, M.-H. Yung, X.-Q. Zhou, P.~J. Love,
  A.~Aspuru-Guzik, and J.~L. O’Brien, ``A variational eigenvalue solver on a
  photonic quantum processor,'' \emph{Nature Communications}, vol.~5, no.~1,
  pp. 1--7, 2014.
\bibitem{jones2013lowToffoliTgate}
C.~Jones, ``Low-overhead constructions for the fault-tolerant {Toffoli} gate,''
  \emph{Physical Review A}, vol.~87, no.~2, p. 022328, 2013.
  
\bibitem{cleve2000parallelQFT}
R.~Cleve and J.~Watrous, ``Fast parallel circuits for the quantum {Fourier}
  transform,'' in \emph{Proceedings of the 41st Annual Symposium on Foundations
  of Computer Science}.\hskip 1em plus 0.5em minus 0.4em\relax IEEE, 2000, pp.
  526--536.






\bibitem{shende2006synthesis3}
V.~V. Shende, S.~S. Bullock, and I.~L. Markov, ``Synthesis of quantum-logic
  circuits,'' \emph{IEEE Transactions on Computer-Aided Design of Integrated
  Circuits and Systems}, vol.~25, no.~6, pp. 1000--1010, 2006.



\bibitem{sivarajah2020tket}
S.~Sivarajah, S.~Dilkes, A.~Cowtan, W.~Simmons, A.~Edgington, and R.~Duncan,
  ``t$|$ket$\rangle$ : A retargetable compiler for {NISQ} devices,''
  \emph{Quantum Science and Technology}, vol.~6, no.~1, p. 014003, 2020.

\bibitem{smith2021leap}
E.~Smith, M.~G. Davis, J.~M. Larson, E.~Younis, C.~Iancu, and W.~Lavrijsen,
  ``{LEAP:} Scaling numerical optimization based synthesis using an incremental
  approach,'' \emph{ACM Transactions on Quantum Computing}, doi:10.1145/3548693.


\bibitem{sundaresan2020echocr}
N.~Sundaresan, I.~Lauer, E.~Pritchett, E.~Magesan, P.~Jurcevic, and J.~M.
  Gambetta, ``Reducing unitary and spectator errors in cross resonance with
  optimized rotary echoes,'' \emph{PRX Quantum}, vol.~1, no.~2, p. 020318,
  2020.


\bibitem{takahashi2009adder}
Y.~Takahashi, S.~Tani, and N.~Kunihiro, ``Quantum addition circuits and
  unbounded fan-out,'' \emph{arXiv:0910.2530}, 2009.

\bibitem{takitaibmheavyhex}
M.~Takita and T.~J. Yoder, ``How {IBM} quantum is advancing quantum error
  correction with hardware experiments,'' 2022.


\bibitem{vartiainen2004synthesis}
J.~J. Vartiainen, M.~M{\"o}tt{\"o}nen, and M.~M. Salomaa, ``Efficient
  decomposition of quantum gates,'' \emph{Physical Review Letters}, vol.~92,
  no.~17, p. 177902, 2004.


\bibitem{wille2007decod24}
R.~Wille and D.~Gro{\ss}e, ``Fast exact {Toffoli} network synthesis of
  reversible logic,'' in \emph{IEEE/ACM International Conference on
  Computer-Aided Design}.\hskip 1em plus 0.5em minus 0.4em\relax IEEE, 2007,
  pp. 60--64.

\bibitem{revlib}
R.~Wille, D.~Große, L.~Teuber, G.~W. Dueck, and R.~Drechsler, ``{RevLib}: An
  online resource for reversible functions and reversible circuits,'' in
  \emph{38th International Symposium on Multiple Valued Logic}, 2008, pp.
  220--225.

\bibitem{younis2021qfast_topologyaware}
E.~Younis, K.~Sen, K.~Yelick, and C.~Iancu, ``{QFAST}: Conflating search and
  numerical optimization for scalable quantum circuit synthesis,'' in
  \emph{International Conference on Quantum Computing and Engineering)}.\hskip
  1em plus 0.5em minus 0.4em\relax IEEE, 2021, pp. 232--243.

\end{thebibliography}
\end{document}